\documentclass[12pt,a4paper]{article}
\usepackage{amsmath}
\usepackage[dvips]{graphicx}
\input epsf
\usepackage{times}
\begin{document}

\begin{titlepage}
\title{Unitarity at the LHC energies}
\author{S. M. Troshin,
 N. E. Tyurin\\[1ex]
\small  \it Institute for High Energy Physics,\\
\small  \it Protvino, Moscow Region, 142281, Russia}
\normalsize
\date{}
\maketitle

\begin{abstract}

Phenomena related to the non-perturbative aspects of strong interactions
 at the LHC  are
discussed with
emphasis
on  elastic and inelastic
soft and hard diffraction processes.
 Predictions for the global characteristics and angular distributions
in
proton-proton collisions with elastic and multiparticle final states
   are given.
 Potential
for  discovery of the novel effects related to the increasing role of the elastic
scattering at the LHC energies and its
physical implications in diffractive  and multiparticle
production processes are reviewed.

\end{abstract}
\end{titlepage}

\section*{Introduction}

The knowledge of hadron structure and hadron interaction dynamics
is the ultimate goal of strong interaction theory.
Nowadays quantum chromodynamics (QCD) is generally accepted as such a theory.
Perturbative QCD enables one to describe successfully spin-averaged observables
 at short distances. However, perturbative calculations can not be applied already at
the distances larger than $0.1$ fm due to chiral symmetry breaking. Moreover,
any real hadron hard interaction process involves also  long range
interaction at some stage.
The fundamental problems of the strong interactions theory are well known and related
to confinement and chiral symmetry breaking phenomena. These phenomena have deal with
collective, coherent properties of quarks and gluons.
Closely related are the problems of the spin structure of a nucleon and
helicity non-conservation on the
quark and hadron levels.

 From the phenomenological point of view all above mentioned phenomena take place in
 the region of diffractive physics.
Thus
understanding of the diffractive interactions plays a fundamental
role under the studies of high-energy limit of QCD.
Studies of diffractive
processes are also important  in the broad context of hadron physics
problems \cite{bjork}.
It was rather surprising that a
high relative probability of coherent processes at high energies
was revealed in the experiments on
hard diffraction at CERN  \cite{cern} and  diffractive
events in the deep-inelastic scattering at HERA \cite{h1,zeus}.
Significant fraction of high-$t$ events among the diffractive
events  in deep-inelastic scattering and in hadron-hadron
interactions were also observed at HERA \cite{herdif} and Tevatron
\cite{tevdif} respectively. These experimental results
renewed interest in the further experimental and theoretical
  studies of   diffractive
 processes and stimulated interest to the
 dedicated QCD experimental
studies at the LHC
(cf. \cite{felix}).

Among the  global problems of strong interactions  the total
cross--section behaviour and its rise constitute a most important question. There are
various approaches which provide the total cross-section increase
with energy but the reason leading to
such behaviour  remains obscure. The nature
of the total--cross section rising energy dependence currently is not understood
since underlying microscopic mechanism is dominated by
 the non-perturbative QCD effects and even rather simple question on universality
 or non-universality of this mechanism has no definite answer.
An important role here belongs to elastic scattering where hadron constituents
interact coherently.
Single diffraction dissociation is a
most simple inelastic diffractive process and studies of soft and
hard final states in this process could become the next
 step after the
elastic scattering.

Multiparticle production and the global  observables such as
mean multiplicity and its energy dependence alongside
with the total, elastic and inelastic cross--sections  provide a clue
to the mechanisms of confinement and hadronization.

General principles  play an essential role in the nonperturbative sector of QCD and
unitarity which regulates the relative strength
of elastic and inelastic processes is  the most important one.
Owing to the experimental efforts during
  recent decades it has become evident that coherent elastic scattering
process will  survive at  high energies and in particular at the
LHC.
However, it is not evident: will hadron
interaction remain to be dominated by multiparticle production?
Or at some distances elastic scattering channel can become playing
 a dominating role at the LHC energy.
This question constitutes  an important problem
 for the background estimates
for the LHC experiments.
We give here  arguments in favor of the
 unorthodox point of view, i.e.
we would like to discuss  possible realization of the new scattering mode where
elastic scattering prevails at
super high energies
and consider   experimental signatures in the
studies of hadronic interactions
at the LHC. Such experiments will be crucial for
understanding of  microscopic nature of  the driving mechanism which provide
rising cross--section, its possible parton structure, high-energy limit
of strong interactions and  approach to  the asymptotical region.

Prevalent role of elastic scattering at very high
energies   has in some extent been implied
  by the limitations
for inelastic  processes obtained on the basis of the general
principles. In particular, it has been shown that the effective interaction
radius of any inelastic process cannot be greater than the interaction
radius of the corresponding (i.e. the process with the same particles in the
initial state) elastic scattering process \cite{logng}.

The appearance of antishadowing
would be associated with significant spin correlations of the
produced particles. We briefly mention some spin related experimental
possibilities then.

In general we  review here particular problems in hadron interactions
which in some cases closely or in other cases
not that much but related to the phenomena of antishadowing.
Several original results  have already
been published in \cite{reltot}, others are
discussed here for the first time.

\section{Approach to asymptotical region and\\ unitarization methods}
It is always important to know how far the asymptotical region lie.
Unfortunately,  at the moment the answer for the above question
can be given in the  model dependent way only and currently there is no
an universal criterion. There are many model parameterizations for
the total cross-sections which use $\ln^2 s$ dependence for
$\sigma_{tot}(s)$. Such models were used widely since
the first CERN ISR results had appeared\footnote{
The first model which provide $\ln^2 s$ dependence for total cross--section
was developed by Heisenberg \cite{heis}.}. This implies the saturation of the
Froissart--Martin bound, however with coefficient in front of $\ln^2 s$ which
 is lower than the asymptotical bound
\footnote{It is known but not often mentioned fact, that the amplitude
which provides an exact saturation
 of the Froissart--Martin bound does correspond to pure elastic scattering.}.
On the other
side  the power-like parameterizations of $\sigma_{tot}(s)$
disregard the Froissart--Martin bound and considers it as a matter of the
 very distant asymptopia.  Both  approaches provide successful fits to the
  experimental data
 in the  available  energy range and even lead to similar predictions for the
 LHC energies.

 However,
  it is not clear whether the  power--like energy dependence would
  obey unitarity bound for the partial--wave amplitudes
   at the LHC energies and beyond.
Meanwhile as it is mentioned unitarity is an important principle which is
 needed to be fulfilled anyway.
 The most straightforward  way is to construct an amplitude which
  \it{ab initio} \rm   satisfy unitarity. But the most
 common way consists in use of
an unitarization
procedure of  some input power-like ``amplitude''.
Unitarization provides a complicated energy dependence of $\sigma_{tot}(s)$
 which can be
approximated by the various functional forms depending on
particular energy range under consideration. These forms providing
a good description of the experimental data in the limited energy
range  have nothing to do with the true asymptotical dependence
$\ln ^2 s$. Of course,  unitarization will lead to  the $\ln ^2s$
dependence but only at $s\to\infty$.

 Unitarity of the scattering matrix $SS^+=1$ implies, in principle, an
existence at high energies $s>s_0$, where $s_0$ is some threshold,
 of the new scattering mode --
antishadow one. It has been revealed in \cite{bbla} and
 described in some detail (cf.
\cite{ech} and references therein) and the most important feature
of this mode is the self-damping of the  contribution from the
inelastic channels.

We argue here that
  the antishadow scattering mode could be definitely
revealed at  the LHC energies and
 give quantitative and qualitative predictions based on the rational unitarization,
i.e. $U$-matrix
unitarization method
 \cite{ltkhs}. There is no  universal, generally accepted
method to implement unitarity
in high energy scattering and as a result of this fact a related problem of the absorptive
corrections role and their sign
has a long history (cf. \cite{sachbla} and references therein).
However, a choice of particular unitarization scheme is not just a matter
of taste. Long time ago the arguments based on analytical properties of the scattering
amplitude  were put forward \cite{blan} in favor of the rational form of unitarization.
It was shown  that this form of unitarization reproduced
correct analytical properties of the scattering amplitude
in the complex energy plane much easier compared to the
exponential form, (simple eikonal singularities
would lead to an essential singularities in the amplitude).
 In potential scattering the eikonal (exponential)
and $U$--matrix (rational)
forms of unitarization correspond to two different approximations
of the scattering wave function, which satisfy
the Schr\"odinger equation to the same order. Rational form of unitarization
corresponds to an approximate wave function which changes both
the phase and amplitude of the wave. This form follows from dispersion
theory. It can be rewritten in the exponential
form but with completely different resultant phase function, and
relation of the two phase functions is given in \cite{blan}.

\section{Unitarity: particle production and\\  elastic scattering}
In the impact parameter representation the unitarity relation
written for the elastic scattering amplitude $f(s,b)$ at high
energies has the form
\begin{equation}
Im f(s,b)=|f(s,b)|^2+\eta(s,b) \label{unt}
\end{equation}
where the inelastic overlap function $\eta(s,b)$ is the sum of
all inelastic channel contributions.  It can be expressed as
a sum of $n$--particle production cross--sections at the
given impact parameter
\begin{equation}
\eta(s,b)=\sum_n\sigma_n(s,b).
\end{equation}
The impact parameter $b$ has a simple geometrical meaning as the
distance in the transverse plane between the centers of the two
colliding hadrons.
 Unitarity equation  has the
two solutions for the case of pure imaginary amplitude:
\begin{equation}
f(s,b)=\frac{i}{2}[1-\sqrt{1-4\eta(s,b)}],\label{usol}
\end{equation}
\begin{equation}
f(s,b)=\frac{i}{2}[1+\sqrt{1-4\eta(s,b)}].\label{usolp}
\end{equation}
Almost everywhere  the second solution is  not taken into account,
since $f(s,b)\to 0$ and $\eta(s,b)\to 0$ at $b\to\infty$.
However, there is nothing wrong with the second solution in the limited
region of impact parameters. Existence of the second solution
leads to interesting experimental predictions and
should be taken into account. Both solutions of unitarity
are naturally reproduced by the rational ($U$--matrix) form
of unitarization.
In the $U$--matrix approach
the form of the elastic scattering amplitude in the
impact parameter representation
is the following:
\begin{equation}
f(s,b)=\frac{U(s,b)}{1-iU(s,b)}. \label{um}
\end{equation}
 $U(s,b)$ is the generalized reaction matrix, which is considered to be an
input dynamical quantity similar to eikonal function. It is worth noting
that transition to antishadowing at small impact parameters
 can be incorporated into eikonal unitarization,
however, the latter should have a very peculiar form.
Inelastic overlap function
is connected with $U(s,b)$ by the relation
\begin{equation}
\eta(s,b)=\frac{\mbox{Im} U(s,b)}{|1-iU(s,b)|^{2}}\label{uf}.
\end{equation}

 Construction of the  particular models in the framework of the $U$--matrix
approach proceeds the common steps, i.e. the basic dynamics as
well as the notions on hadron structure being used  to obtain a
particular form for the $U$--matrix. $U$--matrix
unitarization scheme and  eikonal scheme
 lead to
different predictions for the asymptotical behaviour of
inelastic cross--section and for the ratio of elastic to total
cross-section. This ratio in the $U$--matrix unitarization scheme
reaches its maximal possible value at $s\rightarrow \infty$, i.e.
growth of the elastic cross--section is to be steeper than the growth
of the inelastic cross--section beyond some threshold energy
\begin{equation}\label{umat}
\sigma_{el}(s)\sim\sigma_{tot}(s)\sim\ln^2 s,\quad
\sigma_{inel}(s)\sim\ln s,
\end{equation}
which reflects in fact that the upper bound for the partial--wave
 amplitude in the $U$--matrix
approach (unitarity limit) is $|f_l|\leq 1$ while  the bound for the
case of imaginary eikonal is (black disk limit): $|f_l|\leq 1/2$.
When the amplitude exceeds the black disk limit (in central
collisions at high energies) then the scattering at such impact
parameters turns out to be of an  antishadow nature.
 In this antishadow scattering mode
 the elastic amplitude increases with decrease of the inelastic
 channels contribution.

It is worth noting that the shadow scattering mode is  considered
usually as the only possible one. But as it was already mentioned
existence of the second solution of
unitarity in the limited range of impact parameters is completely
lawful and an antishadow scattering mode
 should not be excluded. Antishadowing can occur
 in the limited region of impact parameters $b<R(s)$ (while
 at large impact parameters only shadow scattering mode can be
 realized. Shadow scattering mode can exist without antishadowing,
 but the opposite statement is not valid.

 Appearance of the antishadow scattering mode
is consistent with the basic idea that the particle
production is the driving force for elastic scattering. Indeed,
the imaginary part of the generalized reaction matrix is the sum
of inelastic channel
 contributions:
\begin{equation}
\mbox{Im} U(s,b)=\sum_n \bar{U}_n(s,b),\label{vvv}
\end{equation}
where $n$ runs over all inelastic states and
\begin{equation}
\bar{U}_n(s,b)=\int d\Gamma_n |U_n(s,b,\{\zeta_n\}|^2
\end{equation}
and $d\Gamma_n$ is the $n$--particle element of the phase space
volume.
The functions $U_n(s,b,\{\zeta_n\})$ are determined by the dynamics
 of $2\rightarrow n$ processes. Thus, the quantity $\mbox{Im}U(s,b)$ itself
 is a shadow of the inelastic processes.
However, unitarity leads to  self--damping of the inelastic
channels \cite{bbl} and increase of the function $\mbox{Im}U(s,b)$ results in
decrease
 of the inelastic overlap function $\eta(s,b)$ when $\mbox{Im}U(s,b)$ exceeds unity.

Respective inclusive cross--section
\cite{tmf,gluas} which takes into account unitarity in the direct channel
  has the form \begin{equation}
\frac{d\sigma}{d\zeta}= 8\pi\int_0^\infty
bdb\frac{I(s,b,\zeta)} {|1-iU(s,b)|^2}.\label{un}
\end{equation}
The function $I(s,b,\zeta)$ in Eq. (\ref{un}) is expressed via the functions
$U_n (s,b,\zeta,\{\zeta _{n-1}\})$  determined  by the dynamics
of the processes  $h_1+h_2\rightarrow
h_3+X_{n-1}$:
\begin{equation}\label{idef}
I(s,b,\zeta)=\sum_{n\geq 3}n\int d\Gamma_n |U_n (s,b,\zeta,\{\zeta _{n-1}\})|^2
\end{equation}
and
\begin{equation}\label{sr}
\int I(s,b,\zeta )d\zeta =\bar n(s,b) \mbox{Im} U(s,b).
\end{equation}

 The kinematical variables $\zeta$ ($x$ and $p_\perp$,
 for example) refer to the produced particle $h_3$ and
 the set of variables $\{\zeta_{n-1}\}$ describe the system $X_{n-1}$
 of $n-1$ particles.

Let us consider now  transition to the antishadow scattering mode, which was
revealed in \cite{bbla}. With
conventional parameterizations of the $U$--matrix (which provide rising cross--sections)
inelastic overlap function increases with energies
at modest values of $s$. It reaches its maximum value $\eta(s,b=0)=1/4$ at some
energy $s=s_0$ and beyond this energy the  antishadow
scattering starts  to develop at small values of $b$ first. The region of energies and
impact parameters corresponding
to the antishadow scattering mode is determined by the conditions
$\mbox{Im} f(s,b)> 1/2$ and $\eta(s,b)< 1/4$.
The quantitative analysis of the experimental data
 \cite{pras} gives the threshold value of energy: $\sqrt{s_0}\simeq 2$ TeV.
This value is confirmed by the another model considerations
\cite{laslo}.

Thus, the function $\eta(s,b)$ becomes peripheral when energy is increasing
beyond $s=s_0$.
At such energies the inelastic overlap function reaches its maximum
 value at $b=R(s)$ where $R(s)$ is the interaction radius.
So, beyond the transition energy there are two regions in impact
 parameter space: the central region
of antishadow scattering at $b< R(s)$ and the peripheral region of
shadow scattering at $b> R(s)$. The impact parameter
dependence of the inelastic channel contribution
$\eta(s,b)$ at $s>s_0$ are represented in Fig. 1 for the case of
standard unitarization scheme  and for the unitarization scheme with
anishadowing.
\begin{center}
\begin{figure}[hbt]
\hspace*{2cm}\epsfxsize= 100  mm  \epsfbox{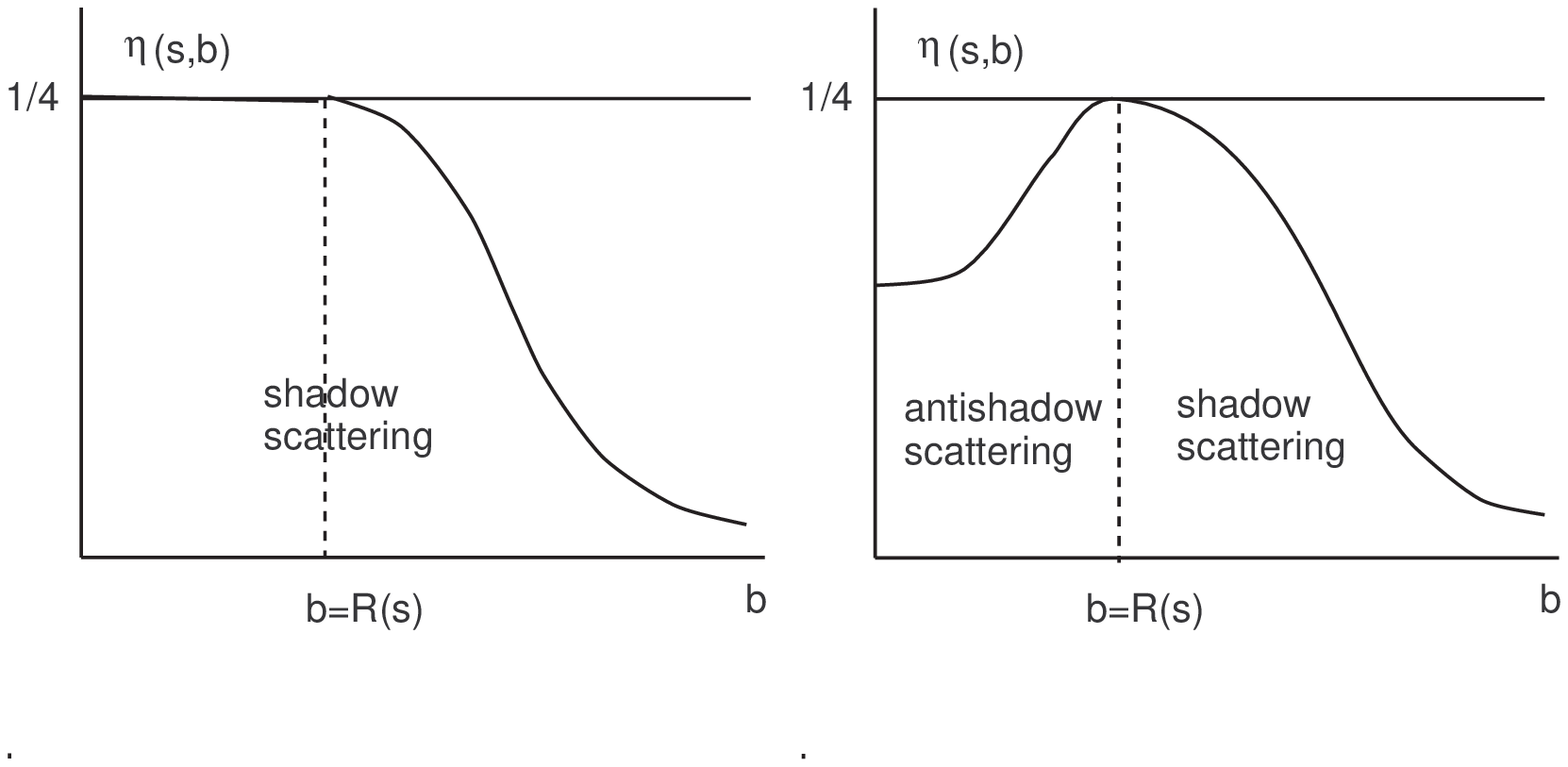}
 \vspace{-1cm}
 \caption{Impact parameter dependence of the inelastic overlap function
 in the standard unitarization scheme (left panel) and in the unitarization scheme
 with antishadowing (right panel).}
 \end{figure}
\end{center}
The region of the LHC energies is the one where antishadow scattering
 mode is to be presented. It will be demonstrated in the next
 section that this mode can be revealed directly measuring
 $\sigma_{el}(s)$ and $\sigma_{tot}(s)$ and not only through the
 analysis in impact parameter representation.

 \section{Approach to asymptotics in the $U$--matrix model}
To get the numerical estimates we shall use the following ansatz
for the generalized reaction matrix
\begin{equation} U(s,b) = ig\left [1+\alpha
\frac{\sqrt{s}}{m_Q}\right]^N \exp(-Mb/\xi )\equiv
ig(s)\exp(-Mb/\xi ), \label{x}
\end{equation} where $M =\sum^N_{q=1}m_Q$.
Here $m_Q$ is the mass of constituent quark, which is
taken to be $0.35$ $GeV$, $N$ is the total number of valence quarks
in the colliding hadrons, i.e. $N=6$ for $pp$--scattering.
The value for the other parameters were obtained in \cite{pras}
and have the following values $g=0.24$, $\xi=2.5$,
$\alpha=0.56\cdot 10^{-4}$.
With these values of parameters the model provides satisfactory
description of the available experimental data for the forward
elastic $pp$-scattering.
To obtain the above explicit form for the function $U(s,b)$
we used chiral quark model for $U$--matrix \cite{chpr}
where $U(s,b)$  is
chosen as a product of the averaged quark amplitudes \begin{equation}
U(s,b) = \prod^{N}_{Q=1} \langle f_Q(s,b)\rangle \end{equation} in
accordance  with the assumed quasi-independence   of  valence
quark scattering.
The $b$--dependence of the function $\langle f_Q \rangle$  has a simple form $\langle
f_Q\rangle\propto\exp(-m_Qb/\xi )$ which correspond to quark interaction radius
$r_Q=\xi/m_Q$.

For the LHC energy $\sqrt{s}= 14$ $TeV$ we have
\begin{equation}\label{s}
 \sigma_{tot}\simeq 230\; \mbox{mb}
\end{equation}
and
\begin{equation}\label{r}
\sigma_{el}/\sigma_{tot}\simeq 0.67.
\end{equation}
\begin{figure}[h]
\begin{center}
  \resizebox{5cm}{!}{\includegraphics*{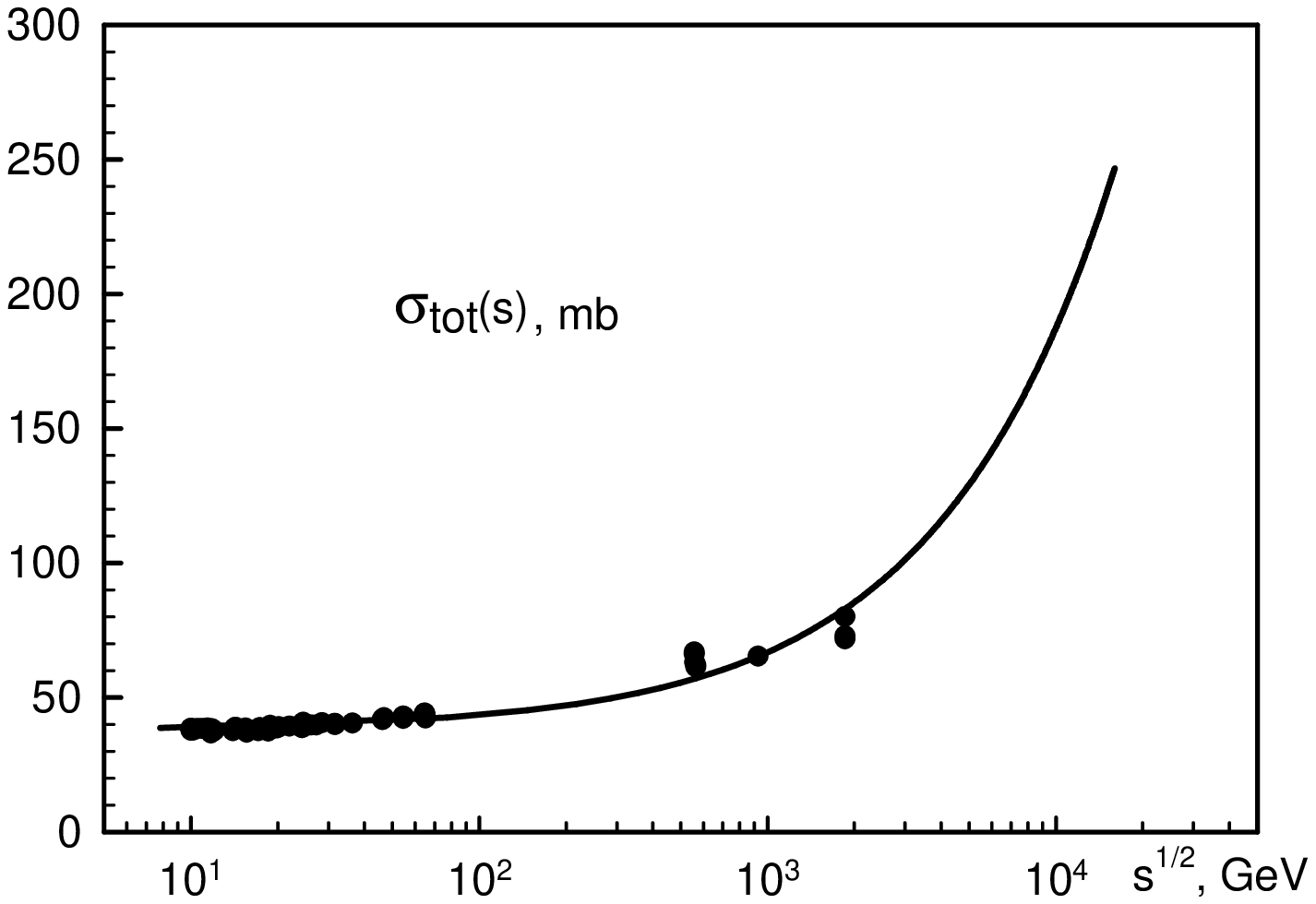}}\;\;\quad
  \resizebox{5cm}{!}{\includegraphics*{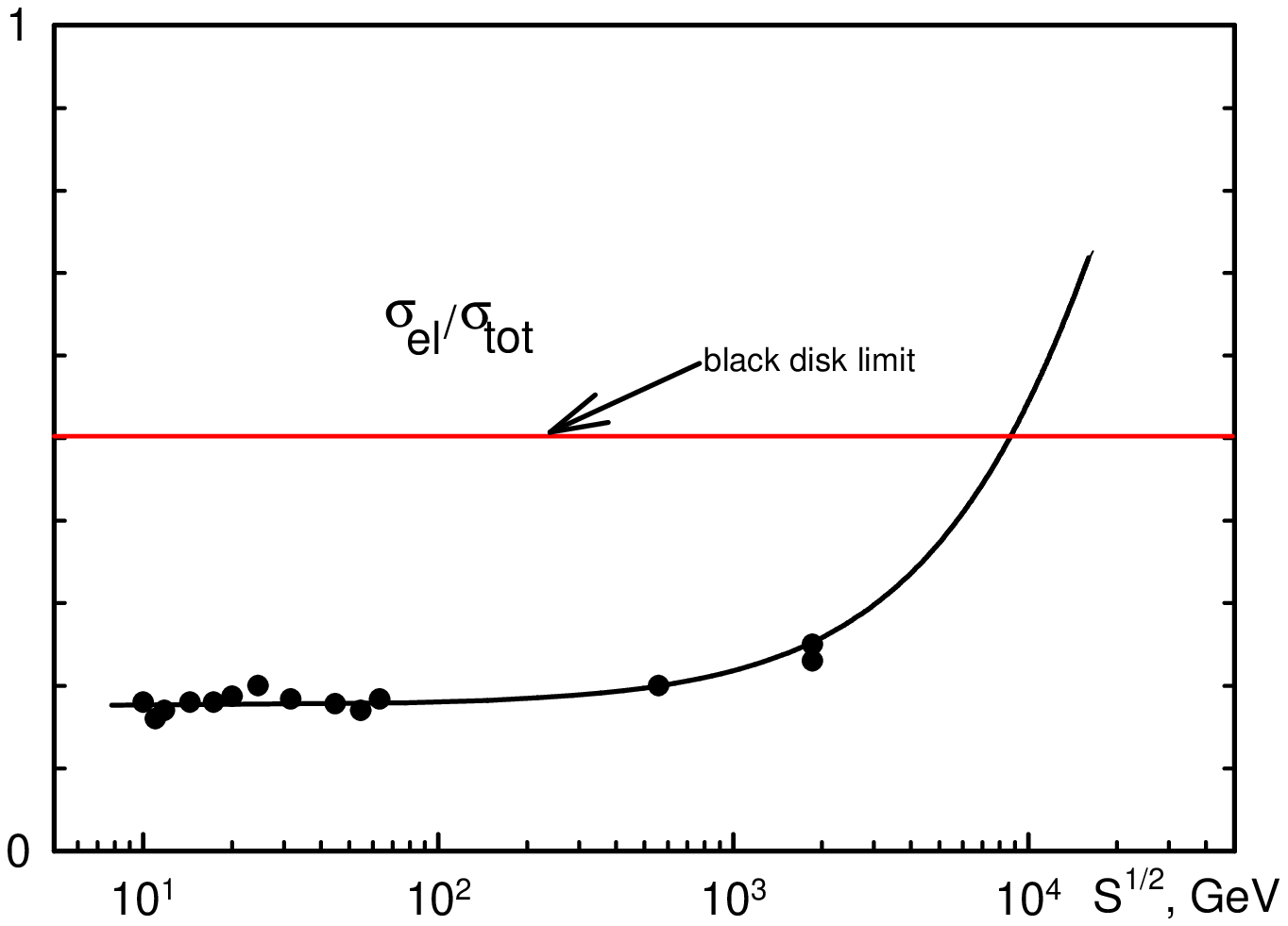}}
\end{center}
\caption{Total and ratio of elastic to total cross-sections of $pp$--interactions}
\label{ts}
\end{figure}
 Thus, the antishadow scattering mode could be discovered
at LHC by measuring $\sigma_{el}/\sigma_{tot}$ ratio which is
greater than the black disc value $1/2$ (cf. Fig. \ref{ts}).

However, the LHC energy is not in the asymptotic region yet; the
total, elastic and inelastic cross-sections behave like
\begin{equation}\label{tot}
  \sigma_{tot,el}\propto \ln^2\left[g\left(1+\alpha
\frac{\sqrt{s}}{m_Q}\right)^N\right],\;
\sigma_{inel}\propto \ln\left[g\left(1+\alpha
\frac{\sqrt{s}}{m_Q}\right)^N\right].
\end{equation}
Asymptotical behavior
\begin{equation}\label{tota}
  \sigma_{tot,el}\propto \ln^2 s,\;\;
 \sigma_{inel}\propto \ln s
\end{equation}
is expected at $\sqrt{s}> 100$ $TeV$.

Another predictions of the chiral quark model is decreasing energy
dependence of the the cross-section of the inelastic diffraction
at $s>s_0$. Decrease of diffractive production cross--section at high energies
($s>s_0$) is due to the  fact  that $\eta  (s,b)$
becomes peripheral at  $s > s_0$  and  the whole  picture  corresponds  to
the antishadow scattering at $b < R(s)$ and to the shadow scattering at
$b>R(s)$ where $R(s)$ is the interaction radius:
\begin{equation}
\frac{d\sigma_{diff}}{dM_X^2}\simeq
\frac{8\pi  g^*\xi ^2}{M_X^2} \eta(s,0).
\end{equation}
The parameter $g^*<1$ is the probability of the excitation of a
constituent quark during interaction.
Diffractive production cross--section  has the familiar
$1/M^2$  dependence     which
is related in  this model to the geometrical size  of excited  constituent
quark.

At the LHC energy $\sqrt{s}=14$ $GeV$  the single
diffractive inelastic cross-sections is limited by the value
$\sigma _{diff}(s)\leq 2.4\;\mbox{mb}$.

The above predicted values for the global characteristics of
$pp$ -- interactions at LHC differ from the most common predictions of
the other models. First, the total cross--section is predicted
to be twice as much of  the common predictions in the range 95-120
mb \cite{vels}
 and it even overshoots the existing cosmic ray data. However,
 extracting total proton--proton cross sections from cosmic ray
 experiments is model dependent and  far from straightforward
 (see, e.g. \cite{bl} and references therein). Those experiments
 measure the attenuation lengths of the showers initiated  by the cosmic
 particles in the atmosphere and are sensitive to the model
 dependent parameter called inelasticity. They do not
 provide any information on elastic scattering channel.

\section{Angular structure of elastic scattering and
diffraction dissociation}

Elastic scattering amplitude $F(s,t)$ is determined by the singularities
in the impact parameter complex plane $\beta=b^2$ plane. It has
poles which positions are determined by the solutions of the following
equation:
\begin{equation}\label{pol}
1+U(s,\beta)=0
\end{equation}
and the branching point at $\beta=0$ (cf. Fig. 3), i.e.
\[
F(s,t)=F_p(s,t)+F_c(s,t).
\]
\begin{figure}[h]
 \begin{center}
\resizebox{8cm}{!}{\includegraphics*{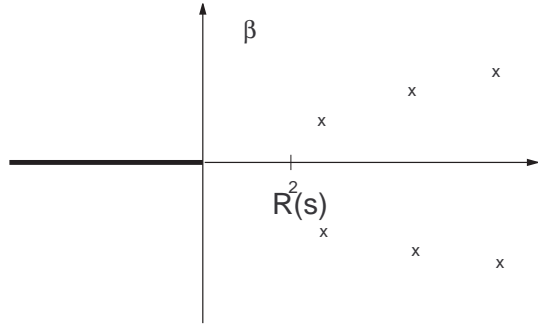}}
\end{center}
\label{poles}
\caption{Singularities of scattering amplitude in the complex $\beta$-plane}
 \end{figure}

Contribution of
the poles located at the points
\[
\beta_n(s)=\left[R(s)+i\frac{\xi}{M}\pi n\right]^2,\quad n=\pm 1,\pm 3,...
\]
where
\[
R(s)=\frac{\xi}{M}\ln g(s)
\]
determines the
 elastic amplitude in the region $|t|/s \ll 1\,(t\neq 0)$.  The
amplitude in this region can be represented in a form of  series
 over the parameter $\tau (\sqrt{-t})$:
\begin{equation}
F(s,t)=s\sum_{k=1}^\infty  \tau ^k(\sqrt{-t})\varphi
_k[R(s),\sqrt{-t}], \label{series}
\end{equation}
where the parameter $\tau $
 decreases exponentially with $\sqrt{-t}$:
\[
\tau (\sqrt{-t})=\exp (-\frac{2\pi\xi }{M }\sqrt{-t}).
\]

This series provides diffraction peak and  dip-bump
structure of the differential cross-section in elastic scattering.
In the region of moderate $t$ it is sufficient to keep
 few or even one of the terms of series Eq. (\ref{series}).
The differential cross-section in this region has well known Orear
behavior.
For elastic scattering amplitude $F(s,t)$ the pole and
cut contributions are decoupled dynamically when $g(s)\rightarrow
\infty  $ at $s\rightarrow \infty$.
At large momentum transfer ($s\to\infty$, $|t|/s$ - fixed) the contribution
from the branching point ($\beta=0$) is a dominating one. The angular distribution
in this region has the power dependence
\begin{equation}\label{pow}
\frac{d\sigma}{dt}\propto \left(\frac{1}{s}\right)^{N+3}f(\theta).
\end{equation}
 There is a direct interrelation of the power
law behavior of the differential cross sections of large angle
scattering with rising behavior of the total
cross sections at high energies.

Similarity between elastic and inelastic diffraction
in the $t$-channel approach suggests that the latter one would
have similar to elastic scattering behavior of the differential
cross-section. However, it cannot be taken for granted and   e.g.
transverse momentum distribution of diffractive events in the
deep-inelastic scattering at HERA  shows a power-like behavior
without apparent dips \cite{will}. Similar behavior was observed
also in the hadronic diffraction dissociation process at CERN
\cite{cern} where also no dip and bump structure was observed.
Angular dependence of diffraction dissociation together with the
measurements of the differential cross--section in elastic
scattering would allow to determine the geometrical properties of
elastic and inelastic diffraction, their similar and distinctive
features and  origin.
In the $U$-matrix approach the
  impact parameter amplitude of diffraction dissociation
$F_{diff}(s,b,M_X)$ can be written in the pure imaginary case as a
square root of the cross-section, i.e.
\begin{equation}\label{ampl}
  F_{diff}(s,b,M_X)={\sqrt{U_{diff}(s,b,M_X)}}/{[1+U(s,b)]}
\end{equation}
and the amplitude $F_{diff}(s,t,M_X)$ is
\begin{equation}\label{amplt}
  F_{diff}(s,t,M_X)=\frac{is}{\pi^2}\int_0^\infty
 bdb J_0(b\sqrt{-t}){\sqrt{U_{diff}(s,b,M_X)}}/{[1+U(s,b)]}.
\end{equation}
The corresponding amplitude $F_{diff}(s,t,M_X)$ can be calculated
analytically. To do so
 we continue the amplitudes
$F_{diff}(s,\beta, M_X),\,\beta =b^2$, to the complex
 $\beta $--plane and then  $F_{diff}(s,t,M_X)$  can be represented as
a sum of the pole contribution and the contribution of the cut \cite{angd}:
\begin{equation}
F_{diff}(s,t, M_X)=F_{diff,p}(s,t, M_X)+F_{diff,c}(s,t, M_X)\label{sum}
\end{equation}

 The situation is different in the case of diffraction
production. Instead of dynamical separation of the pole and cut
contribution discussed above we have a  suppression of the pole
contribution at high energies since  at fixed $t$
\begin{equation}
F_{diff,p}=O(s[g(s)]^{-\frac {M_X}{2{M}}}\ln ^{1/2}
g(s)),\quad F_{diff,c}=O(s[g(s)]^{-\frac {1}{2}}).
\end{equation}
 Therefore, at
all $t$ values   we will have
\begin{equation}
F_{diff}(s,t, M_X)\simeq F_{diff,c}(s,t, M_X),\label{dom}
\end{equation}
where
\begin{equation}
 F_{diff,c}(s,t, M_X)\simeq ig^*g^{-1/2}(s)(1-\frac{t}{\bar M_X^2})^{-3/2},
\end{equation}
where $\bar M_X=(M_X-M-1)/2\xi$. This means that the
differential cross-section of the diffraction production will have
smooth dependence on $t$ with no apparent dips and bumps
\begin{equation}\label{dsig}
\frac {d\sigma_{diff}}{dtdM_X^2}\propto (1-\frac{t}{\bar M_X^2})^{-3}.
\end{equation}
It is interesting to note that at large values of $M_X\gg  M$
the normalized differential cross-section
$\frac{1}{\sigma_0}\frac{d\sigma }{ dtdM_X^2}$ ($\sigma_0$ is the
value of cross-section at $t=0$) will exhibit scaling behavior
\begin{equation}\label{scal}
\frac{1}{\sigma_0}\frac{d\sigma }{dtdM_X^2}=f(-t/M_X^2),
\end{equation}
and explicit form of the function $f(-t/M_X^2)$ is the following
\begin{equation}\label{ftau}
 f(-t/M_X^2)=(1-4\xi ^2t/M_X^2)^{-3}.
\end{equation}
This dependence is depicted on Fig. 4.

\begin{figure}[h]
 \begin{center}
\resizebox{6cm}{!}{\includegraphics*{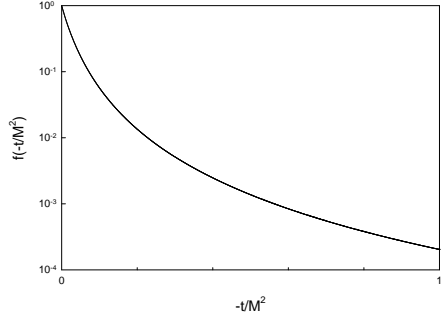}}
\end{center}
\label{scalfig}
\caption{Scaling behavior of the normalized differential cross-section
$\frac{1}{\sigma_0}\frac{d\sigma }{ dtdM^2}$.}
 \end{figure}

 The above scaling has been obtained in the model approach,
however it might have a more general meaning.

 The angular structure of
diffraction dissociation processes given by Eq. (\ref{dsig}) takes
place  at high energies where $g(s)>1$ while at moderate and low energies
where $g(s)\leq 1$ the both contributions from poles and cut are
significant. In this region
\begin{equation}
F_{diff}(s,t, M_X^2)=s\sum_{k=1}^\infty  \tau ^k(\sqrt{-t})\varphi
_{diff,k}[R(s),\sqrt{-t}, M_X^2], \label{poldif}
\end{equation}
where the parameter $\tau(\sqrt{-t})$ is the same as it is in the
elastic scattering. Thus at low energies the situation is
similar to the elastic scattering, i.e.  diffraction cone
and possible dip-bump structure should be present in the region of
small values
 of $t$, and the overall behaviour of the differential
cross-section will be rather complicated and incorporates
diffraction cone, Orear type and power-like dependencies.

However, at high energies a simple power-like dependence on $t$
Eq. (\ref{dsig}) is
predicted. It was shown that the normalized differential
cross-section has a scaling form and only depends  on the ratio
$-t/M_X^2$ at large values of $M_X^2$.

In fact, our particular comparative analysis of the poles and cut
contributions has very little with the  model form of the
$U$--matrix.  This is
why it has a more general meaning.

 At
the LHC energies the diffractive events with the masses as large as
3 TeV could be studied. It would be interesting to check this
prediction at the LHC  where the scaling and simple power-like
behavior of diffraction dissociation differential cross-section
should be observed. Observation of such behavior would confirm the
diffraction mechanism based on excitation of the complex
hadronlike  object -- the constituent quark.

\section{Angular distributions of leading protons  in central production processes}
It was proposed in \cite{petr} to study influence of centrally produced particles
on the angular distribution of leading protons in the processes with two rapidity gaps,
which are also known as a double Pomeron exchange (dpe) processes:
\begin{equation}\label{dpe}
p+p\to p+X+p.
\end{equation}
In what follows we will consider symmetrical case $t_1=t_2=t$.
It is interesting to study
how the diffractive pattern observed in elastic scattering will be changed in the
corresponding processes with centrally produced particles.

First of all elastic scattering and process (\ref{dpe}) are very different
from the point of view of unitarity equation (\ref{unt}). The process (\ref{dpe})
is one of the many contributing processes to the function $\eta(s,b)$ and consequently
into the amplitude of elastic scattering $f(s,b)$.
In the $U$-matrix approach the
  impact parameter amplitude of the process (\ref{dpe})
$F_{dpe}(s,b,\zeta)$ can be written in the pure imaginary case according to
(\ref{un}) as a
square root of the cross-section, i.e.
\begin{equation}\label{ampldpe}
  F_{dpe}(s,b,\zeta)={\sqrt{I(s,b,\zeta)}}/{[1+U(s,b)]}
\end{equation}
and the amplitude $F_{dpe}(s,t,\zeta)$ is
\begin{equation}\label{ampltdpe}
  F_{dpe}(s,t,\zeta)=\frac{is}{\pi^2}\int_0^\infty
 bdb J_0(b\sqrt{-t}){\sqrt{I(s,b,\zeta)}}/{[1+U(s,b)]}.
\end{equation}
The variable $t$ is the momentum transfer to one of the protons, while
the variable $\zeta$ is related to the system of particles $X$ or to one
particle from this system.
Using relation (\ref{sr}) we can represent $I(s,b,\zeta)$ in the form
\begin{equation}\label{irep}
I(s,b,\zeta)=\Phi(s,b,\zeta)\mbox{Im}U(s,b),
\end{equation}
where
\begin{equation}\label{fisr}
\int \Phi(s,b,\zeta) d\zeta =\bar n(s,b).
\end{equation}
For the mean multiplicity we suppose that the multiplicity
of the centrally produced particles is given by the following expression
\begin{equation}\label{mmultcp}
\bar n (s,b)= \beta N_0(s)D_C(b),
\end{equation}
where the function  $D_C(b)$ describes distribution of two hadron condensates
clouds in the overlapping region.
Arguments  in favor of such form and assumed hadron
structure are desribed in the next section.
 The corresponding amplitude
$F_{dpe}(s,t,\zeta)$ can be calculated analytically and calculations are similar
to the case of elastic scattering amplitude and amplitude of diffraction dissociation.
It is necessary
 to continue the amplitudes
$F_{dpe}(s,\beta, \zeta)$ ($\beta =b^2$), to the complex
 $\beta $--plane and transform the Fourier--Bessel integral over\\ impact
parameter into the integral in the complex $\beta $ -- plane over
the contour $C$ which goes around the positive semiaxis.
The amplitude $F_{dpe}(s,\beta , \zeta )$ has the poles and a
  branching point at $\beta =0$.
Therefore it  can be represented as
a sum of the pole contributions and the contribution of the cut:
\begin{equation}
F_{dpe}(s,t, \zeta)=F_{dpe,p}(s,t, \zeta)+F_{dpe,c}(s,t, \zeta)\label{sumdpe}
\end{equation}

Then, using relation (\ref{fisr}) and assuming
\[
\Phi(s,b,\zeta)  =\bar n(s,b)\phi(\zeta)
\]
we obtain that
\begin{equation}\label{poldpe}
F_{dpe,p}\sim s\sum_{n=\pm 1,\pm 3,...}[\bar n(s,\sqrt{\beta _n})\phi(\zeta)]
\sqrt{\beta _n}K_0(\sqrt{t\beta _n}),
\end{equation}
i.e. diffractive pattern of leading protons will depend on the
distribution of mean multiplicity in impact parameter
of the centrally produced particles.
Using for the mean multiplicity results described in Section 6, the amplitude
(\ref{poldpe}) can be rewritten in the form
\begin{equation}\label{poldpen}
F_{dpe,p}\sim s s^{\frac{1}{4}(1-\frac{M_C\xi}{m_Q})}\phi(\zeta)\sum_{n=\pm 1,\pm 3,...}
e^{i\pi n\frac{M_C\xi}{2M}}
\sqrt{\beta _n}K_0(\sqrt{t\beta _n})
\end{equation}
Thus, the presence of oscillating factor $e^{i\pi n\frac{M_C\xi}{2M}}$ would lead to
significant differences in the diffractive patterns of leading protons
in the processes (\ref{dpe}) and elastic scattering. Indeed, at small values of $t$
all terms of the series (\ref{poldpen}) are important. In elastic scattering
the summation over all $n$ leads to the  exponential behavior of the differential
cross section \cite{chpr}:
\begin{equation}\label{elsctdif}
\frac{d\sigma}{dt}\propto \exp\left(B(s)t\right),\quad B(s)\propto \ln ^2 s.
\end{equation}
However, for the process (\ref{dpe}) the  terms with the large
values of $n$ will be suppressed due  the oscillation factor. Thus, we
expect that the Orear type behaviour will lake place already at low values of $t$
and differential
cross section would have the following $t$--dependence already at small and moderate
 values of $t$:
\begin{equation}\label{dpedep}
\frac{d\sigma}{dtd\zeta}\propto \exp\left(-\frac{2\pi\xi}{M}\sqrt{-t}\right).
\end{equation}

\section{Multiparticle production and antishadowing}
 The region of the LHC energies is the one where new,
  antishadow scattering
 mode can be observed.
Immediate  question arises on consistency of the antishadowing with
the growth of mean multiplicity in hadronic collisions with energy.
Moreover, many models and the experimental data suggest a power-like energy dependence
of mean multiplicity\footnote{Recent discussions of the problems
 of multiparticle production processes and rising
mean hadronic multiplicity dependence
can be found in \cite{sissa} and \cite{menon}} and a priori the  compatibility
of such dependence  with  antishadowing is not evident.

Now we turn to the mean multiplicity and consider first the corresponding
quantity in the impact parameter representation. As it follows from (\ref{idef})
and (\ref{sr})
the $n$--particle production
cross--section $\sigma_n(s,b)$
\begin{equation}\label{snb}
\sigma_n(s,b)=\frac{\bar{U}_n(s,b)}{|1-iU(s,b)|^2}
\end{equation}
Then the probability
\begin{equation}\label{pnb}
  P_n(s,b)\equiv\frac{\sigma_n(s,b)}{\sigma_{inel}(s,b)}=\frac{\bar{U}_n(s,b)}{\mbox{Im} U(s,b)}.
\end{equation}

Thus, we observe the cancellation of unitarity corrections in
the ratio of the cross-sections $\sigma_n(s,b)$ and $\sigma_{inel}(s,b)$.
Therefore the mean multiplicity in the impact parameter representation
\[
\bar n (s,b)=\sum_n nP_n(s,b)
\]
 is not affected by unitarity corrections  and
therefore cannot  be proportional
 to $\eta(s,b)$. This conclusion is  consistent with  Eq. (\ref{sr}).
The above mentioned  proportionality is a rather natural assumption in the framework
of the geometrical models, but it is in conflict with
the unitarity. Because of that the results  \cite{enk}
based on such assumption
 should  be taken with precaution.
However, the above cancellation of unitarity corrections
 does not take place for the quantity $\bar n (s)$ which we
 address now.

\begin{figure}[h]
\begin{center}
\resizebox{6cm}{!}{\includegraphics*{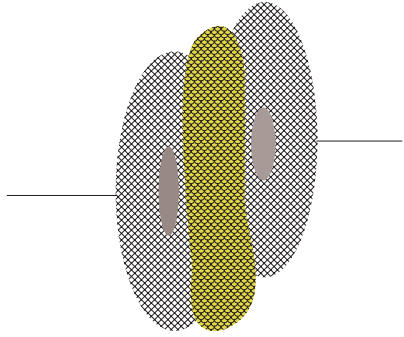}}
\end{center}
 \caption{Schematic view of initial stage of the hadron
 interaction.}
\label{hcol}
\end{figure}

We use a   model for the hadron scattering
 described in \cite{chpr} which
is  based on the ideas of chiral quark models.
The picture of a hadron consisting of constituent quarks embedded
 into quark condensate implies that overlapping and interaction of
peripheral clouds   occur at the first stage of hadron interaction (Fig. \ref{hcol}).
Nonlinear field couplings  could transform then the kinetic energy to
the internal energy and mechanism of such transformation was discussed
 by Heisenberg \cite{heis} and  Carruthers \cite{carr}.
As a result massive
virtual quarks appear in the overlapping region and  some effective
field is generated.
Valence constituent quarks  located in the central part of hadrons are
supposed to scatter simultaneously in a quasi-independent way by this effective
 field.

 Massive virtual quarks play a role
of scatterers for the valence quarks in elastic scattering and
 their hadronization leads to
production of secondary particles in the central region.
 To estimate number
of such scatterers one could assume that  part of hadron energy carried by
the outer condensate clouds is being released in the overlap region
 to generate massive quarks. Then this number can be estimated  by:
 \begin{equation} \tilde{N}(s,b)\,\propto
\,\frac{(1-\langle k_Q\rangle)\sqrt{s}}{m_Q}\;D^{h_1}_c\otimes D^{h_2}_c
\equiv N_0(s)D_C(b),
\label{Nsbt}
\end{equation} where $m_Q$ -- constituent quark mass, $\langle k_Q\rangle $ --
average fraction of
hadron  energy carried  by  the constituent valence quarks. Function $D^h_c$
describes condensate distribution inside the hadron $h$, and $b$ is
an impact parameter of the colliding hadrons.

Thus, $\tilde{N}(s,b)$ quarks appear in addition to $N=n_{h_1}+n_{h_2}$
valence quarks. In elastic scattering those quarks are transient
ones: they are transformed back into the condensates of the final
hadrons. Calculation of elastic scattering amplitude has been performed
in \cite{chpr}.
However,  valence quarks can excite a part of the cloud of the virtual massive
quarks and these virtual massive
 quarks will subsequently fragment into the multiparticle
final states. Such mechanism is responsible for the  particle production
in the fragmentation region and should lead to  strong correlations between
secondary particles. It means that correlations exist
between particles from the same (short--range correlations)
and different clusters (long--range correlations)
 and, in particular, the forward--backward
multiplicity correlations should be observed. This mechanism can be called
 as a correlated cluster
production mechanism. Evidently, similar mechanism should be significantly
reduced in $e^+e^-$--annihilation
processes and therefore large correlations are not to be expected there.

As it was already mentioned simple (not induced by interactions with valence
quarks) hadronization of massive $\tilde{N}(s,b)$
quarks leads to  formation of the multiparticle
final states, i.e. production of the secondary particles in the central region.
The latter should not provide any correlations in the multiplicity
distribution.

Remarkably,  existence of the massive quark-antiquark matter in the stage
preceding
hadronization seems to be
supported  by the experimental data obtained
at CERN SPS and RHIC (see \cite{biro} and references therein).

Since the quarks are constituent, it is natural to expect  direct
proportionality between a secondary particles multiplicity  and
number of virtual massive quarks appeared (due to  both mechanisms of multiparticle
production) in  collision of the  hadrons
with  given impact parameter:
\begin{equation}\label{mmult}
\bar n (s,b)=\alpha (n_{h_1}+n_{h_2})N_0(s)D_F(b)+ \beta N_0(s)D_C(b),
\end{equation}
with  constant factors $\alpha$ and $\beta$ and
\[
D_F(b)\equiv D_Q\otimes D_C,
\]
where the function $D_Q(b)$ is the probability amplitude of the interaction of
valence quark with the excitation of the effective field, which is in fact related
to the quark matter distribution in this hadron-like object called
the valence constituent quark \cite{chpr}.
The mean multiplicity $\bar n(s)$ can be calculated according to the
formula
\begin{equation}\label{mm}
\bar n(s)= \frac{\int_0^\infty  \bar n (s,b)\eta(s,b)bdb}{\int_0^\infty \eta(s,b)bdb}.
\end{equation}
It is evident from Eq. (\ref{mm}) and Fig. 1 that the antishadow
mode with the peripheral profile of $\eta(s,b)$ suppresses the region of small
impact parameters, and  the main contribution to the mean multiplicity is due to
 peripheral region of $b\sim R(s)$.

To make explicit calculations  we model for simplicity
the condensate distribution $D_C(b)$ and the impact parameter dependence
of the probability amplitude $D_Q(b)$
 of the interaction of
valence quark with the excitation of the effective field by the exponential forms,
and thus we use  exponential
 dependencies for the functions $D_F(b)$ and $D_C(b)$  with the
 different radii.
  Then the mean multiplicity
\begin{equation}\label{nsbex}
  \bar n (s,b)=\tilde\alpha N_0(s)\exp (-b/R_F)+\tilde\beta N_0(s)\exp (-b/R_C).
\end{equation}

After calculation of the integrals (\ref{mm})
 we arrive to the power-like dependence
of the mean multiplicity $\bar n(s)$ at high energies
\begin{equation}\label{asm}
\bar n(s) = as^{\delta_F}+bs^{\delta_C},
\end{equation}
where
\[
\delta_{F}={\frac{1}{2}\left(1-\frac{\xi }{m_QR_{F}}\right)}\quad
\mbox{and}\quad \delta_{C}={\frac{1}{2}\left(1-\frac{\xi }{m_QR_{C}}\right)}.
\]
\begin{figure}[h]
\begin{center}
\resizebox{6cm}{!}{\includegraphics*{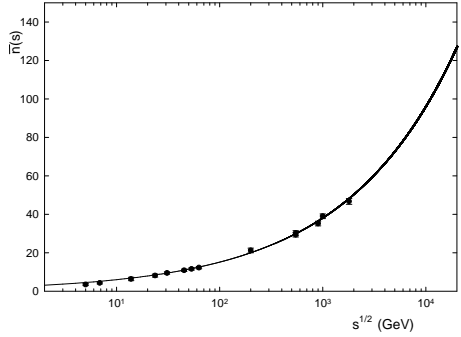}}
\caption{Energy dependence of mean multiplicity, theoretical curve
  is given by the equation $\bar n(s)=as^\delta$ ($a=2.328$, $\delta = 0.201$); experimental
  data from the Refs. \cite{ua5}.}
\end{center}
\label{mean}
\end{figure}
There are four free parameters in the model, $\tilde\alpha$, $\tilde\beta$ and
$R_F$, $R_C$, and the freedom
in their choice is translated to  $a$, $b$ and
 $\delta_F$, $\delta_C$.
 The value of  $\xi=2$ is fixed from the data on angular
 distributions \cite{chpr} and for the mass of constituent quark
 the standard value $m_Q=0.35$ GeV was taken. However, fit to experimental data on the
 mean multiplicity leads to approximate equality $\delta_F\simeq \delta_C$ and
actually Eq. (\ref{asm}) is reduced to the two-parametric power-like energy dependence
 of mean multiplicity
\[
 \bar{n}=as^\delta,
 \]
 which is in good agreement with the experimental data (Fig. \ref{mean}). Equality
 $\delta_F\simeq \delta_C$  means that variation of the correlation strength with
 energy is weaker than the power-like one and could be described, e.g. by
  a logarithmic function of energy.
  From the comparison
 with the data  on mean multiplicity we obtain that
$\delta\simeq 0.2$, which corresponds to the effective masses, which are determined
by the respective radii ($M=1/R$),
$M_C\simeq M_F\simeq 0.3m_Q$, i.e. $M_F\simeq M_C\simeq m_\pi$.

The value of mean multiplicity expected at the LHC energy
($\sqrt{s}=14$ TeV) is about 110.
It is not surprising that it is impossible to differentiate contributions from the
two mechanisms of particle production at the level of mean multiplicity. The studies
of correlations are necessary for that purpose.

Multiplicity distribution $P_n(s,b)$ and mean multiplicity
$\bar n(s,b)$ in the impact parameter representation have
no absorptive corrections, but since
antishadowing leads to suppression of particle
production at small impact parameters and the main contribution to
the integral  multiplicity $\bar n(s)$ comes from
the region of $b\sim R(s)$. Of course, this prediction is to be valid for
the energy range where antishadow scattering mode starts to develop
and is therefore consistent with the ``centrality'' dependence of the mean multiplicity
observed at RHIC \cite{phen}.

It would be interesting to note that due to peripheral
form of the inelastic overlap function the secondary particles will be mainly
 produced at impact parameters $b\sim R(s)$ and thus will carry out large orbital
 angular momentum
\[
L\simeq R(s)\frac{\sqrt{s}}{2}\bar n(s,b=R(s)).
\]
  To compensate this orbital momentum spins of secondary particles should  become
  lined up, i.e. the spins of the produced particles should demonstrate
   significant
  correlations when the antishadow scattering mode appears.
 Similar conclusion on spin correlations of secondary particles has been
 made long time ago by  Yang and Chou  under utilization of the concept of partition
 temperature in the geometric model for hadron production \cite{chou}.

It is also worth  noting  that  no limitations follow from the general principles
 for the mean multiplicity, besides the well known one based
on the energy conservation law.
Having in mind relation (\ref{nsbex}), we could say  that the obtained power--like dependence
which takes into account unitarity effects could be considered as a kind of a saturated
upper bound  for the mean multiplicity growth.

Elastic scattering domination at the LHC and
 the appearance of the antishadow
scattering mode  implies a somewhat unusual
scattering picture. At high energies the proton should be
represented as a very loosely bounded composite system and it
appears that this system has a high probability to reinstate
itself only in the central collisions where all of its parts
participate in the coherent interactions. Therefore the central
collisions are responsible for elastic processes while the
peripheral ones where only few parts of weekly bounded protons are
involved result in the production of the secondary particles. This
leads to the peripheral impact parameter profile of the inelastic
overlap function. Such evolution could be accomplished
 with spin correlations of the produced particles.

\section{Polarization measurements}

In  soft hadronic interactions  significant single-spin
effects could be expected since the helicity conservation
does not work for interactions at large distances,
 once the chiral $SU(3)_L\times SU(3)_R$ symmetry of
 the QCD Lagrangian is spontaneously broken in the real world.
Thus, studies of the $p_{\perp}$--dependence of the one--spin
asymmetries can be used as a way to reveal a transition from a
non--pertur\-ba\-tive phase ($P\neq 0$) to the perturbative
one ($P=0$). The essential point here is an assumption
 that at  short distances the vacuum is perturbative.
However, the very existence of the above transition can
 not be taken
for granted since the vacuum, even at
short distances, could be filled up with the fluctuations of gluon and quark
fields.
The measurements of the one--spin transverse
asymmetries and polarization  is an important probe of the
 chiral structure of the effective QCD Lagrangian.

At the same time we can note that polarization effects
as well as some other recent
experimental data demonstrate
that hadron interactions have a significant degree of coherence.
Experimentally,  spin asymmetries  increase  at
high transverse momentum in elastic scattering and are flat in inclusive
processes.

It is interesting to note that on the base of the model in \cite{pollam}
 one should expect
a zero polarization in the
region where quark--gluon plasma (QGP)  has been  formed,
 since  chiral symmetry is restored
and there is no room for  quasiparticles such as  constituent quarks.
Thus, the absence or strong diminishing e. g. of transverse  hyperon polarization
can be used
 as a signal of QGP formation in heavy-ion collisions.
This prediction should also be valid
for the  models based on confinement, e.g. the Lund and Thomas precession
model.  We could use a vanishing polarization
of e. g. $\Lambda$--hyperons  in heavy ion collisions
 as a sole result of QGP formation provided
the corresponding observable is non-zero  in  proton--proton collisions.
 The prediction based on this observation
would be a decreasing behavior of polarization of $\Lambda$ with the impact parameter
in heavy-ion collisions in the region of energies and densities where QGP was
produced:
\begin{equation}\label{zer}
P_\Lambda(b)\to 0\quad \mbox{at}\quad b\to 0,
\end{equation}
since the overlap is maximal at $b=0$. The value of the impact parameter  can be
controlled by the centrality in heavy--ion collisions.
The experimental program could therefore
include  measurements of $\Lambda$--polarization
in $pp$--interactions first, and then if a significant polarization would be
measured,  the corresponding measurements could be a useful tool for the
 QGP detection.
Such measurements seem to be experimentally feasible
at  RHIC and  LHC provided it is supplemented with forward
detectors.
\section*{Conclusion}
The possibility to reveal a new scattering mode at the LHC is an intriguing one.
It would significantly change our picture of hadron scattering and lead to better
understanding of the non-perturbative region of QCD. Diffraction and related processes
are very important for studies of collective, coherent phenomena in hadronic
 interactions. Predictions for the experimental observables in these
  processes presented in this review and their experimental verifications will certainly
   increase the scope of strong interaction  studies at the LHC.

\section*{Acknowledgement}
We are grateful to V. Petrov and A.~De~Roeck   for useful suggestions,
many interesting discussions of
the results and problems of hadronic interactions at the LHC.

\end{document}